\documentclass[letterpaper,english,prl,reprint]{revtex4-1}
\usepackage{lmodern}

\usepackage[T1]{fontenc}
\usepackage[latin9]{inputenc}
\usepackage{color}
\usepackage{bm}
\usepackage{amsmath}
\usepackage{graphicx}
\usepackage{amssymb}

\makeatletter

 \@ifundefined{textcolor}{}
 {%
   \definecolor{BLACK}{gray}{0}
   \definecolor{WHITE}{gray}{1}
   \definecolor{RED}{rgb}{1,0,0}
   \definecolor{GREEN}{rgb}{0,1,0}
   \definecolor{BLUE}{rgb}{0,0,1}
   \definecolor{CYAN}{cmyk}{1,0,0,0}
   \definecolor{MAGENTA}{cmyk}{0,1,0,0}
   \definecolor{YELLOW}{cmyk}{0,0,1,0}
 }

\usepackage{bm}

\makeatother

\usepackage{babel}

\begin{document}

\title{Transport-driven toroidal rotation in the tokamak edge}

\author{T. \surname{Stoltzfus-Dueck}}

\email{tstoltzf@ipp.mpg.de}

\affiliation{Max-Planck-Institut für Plasmaphysik, Boltzmannstr. 2, D-85748 Garching,
Germany}
\begin{abstract}
The interaction of passing-ion drift orbits with spatially-inhomogeneous
but purely diffusive radial transport is demonstrated to cause spontaneous
toroidal spin-up to experimentally-relevant values in the tokamak
edge. Physically, major-radial orbit shifts cause orbit-averaged diffusivities
to depend on $v_{\parallel}$, including its sign, leading to residual
stress. The resulting intrinsic rotation scales with $T_{i}/B_{\theta}$,
resembling typical experimental scalings. Additionally, an inboard
(outboard) X-point is expected to enhance co- (counter-) current rotation.
\end{abstract}



\maketitle
\newcommand{\xx}{x}
\newcommand{\yy}{y}
\newcommand{\vn}{v}
\newcommand{\xb}{\bar{x}}
\newcommand{\yb}{\bar{y}}
\newcommand{\uu}{u}
\newcommand{\yp}{\yy^{\prime}}
\newcommand{\sx}{\xi}
\newcommand{\yg}{\tau}
\newcommand{\pt}{\partial_{t}}
\newcommand{\px}{\partial_{\xx}}
\newcommand{\py}{\partial_{\yy}}
\newcommand{\pxb}{\partial_{\xb}}
\newcommand{\pyb}{\partial_{\yb}}
\newcommand{\pu}{\partial_{\uu}}
\newcommand{\dd}{d}
\newcommand{\fie}{f_{i}}
\newcommand{\fio}{f_{i0}}
\newcommand{\drat}{\delta}
\newcommand{\yo}{y_{0}}
\newcommand{\dy}{D}
\newcommand{\dyo}{D_{y0}}
\newcommand{\deff}{D_{\text{eff}}}
\newcommand{\dnaught}{D_{0}}
\newcommand{\fb}{d_{c}}
\newcommand{\boldp}[1]{\bm{}}
\newcommand{\boldlp}{\bm{(}}
\newcommand{\boldrp}{\bm{)}}
\newcommand{\fluxs}{\Gamma}
\newcommand{\gf}{G}
\newcommand{\at}{a_{2}}
\newcommand{\aone}{a_{1}}
\newcommand{\fluxp}{\Gamma^{p}}
\newcommand{\fluxm}{\Pi}
\newcommand{\fluxh}{Q_{\parallel}}
\newcommand{\vint}{\vn_{\text{int}}}
\newcommand{\vintd}{\vint^{\text{dim}}}
\newcommand{\vti}{v_{ti}}
\newcommand{\bpo}{B_{\theta}}
\newcommand{\bto}{B_{\phi}}
\newcommand{\bom}{B_{0}}
\newcommand{\ri}{\rho_{i}}
\newcommand{\qq}{q}
\newcommand{\lp}{L_{\perp}}
\newcommand{\Ro}{R_{0}}
\newcommand{\ro}{a}
\newcommand{\Ti}{T_{i}}
\newcommand{\funb}{f_{\text{unb}}}
\newcommand{\fconvi}{f_{c}}
\newcommand{\fnbi}{f_{\text{NBI}}}
\newcommand{\vbeam}{v_{\text{NBI}}}
\newcommand{\vrig}{v_{\text{rig}}}
\newcommand{\Ll}{L_{\parallel}}
\newcommand{\vlf}{\tilde{v}_{\parallel}}
\newcommand{\vrf}{\tilde{v}_{r}}
\newcommand{\kp}{k_{\perp}}
\newcommand{\ver}{\tilde{v}_{E,r}}
\newcommand{\kl}{k_{\parallel}}
\newcommand{\ped}{|_{\text{pt}}}
\newcommand{\sep}{|_{\text{sep}}}
\newcommand{\nni}{n_{i}}
\newcommand{\vl}{v_{\parallel}}
\newcommand{\lf}{L_{\phi}}
\newcommand{\exb}{\mathbf{E}\times\mathbf{B}}
\newcommand{\nii}{\nu_{ii}}
\newcommand{\bv}{\mathbf{B}}
\newcommand{\plascur}{I_{p}}
\newcommand{\logf}{g}
\newcommand{\fif}{\fie^{\left(1\right)}}
\newcommand{\Er}{E_{r}}
\newcommand{\tc}{\tau_{\text{cr}}}
Rotation patterns strongly affected
by turbulent momentum transport are broadly observed in nature, for
example in atmospheric flows, stellar interiors, and accretion disks
\cite{Kichatinov86,*Frisch87,*Balbus98,*Miesch09}. Laboratory tokamak
plasmas are observed to rotate toroidally in the absence of applied
torque, with edge rotation directed with the plasma current (co-current),
often proportional to plasma stored energy $W$ over plasma current
$\plascur$, and reaching tenths of the ion thermal speed $\vti$
\cite{*[{}] [{, and references therein. }] deGrassie09rev,Rice07}.
Such intrinsic rotation is of practical as well as fundamental interest,
since it stabilizes certain instabilities \cite{Strait95} and contributes
to a sheared radial electric field $E_{r}$, believed to suppress
turbulent transport \cite{Biglari90}. Intrinsic rotation is of special
importance for the next-generation tokamak ITER, since $\alpha$-heating
(nuclear fusion) applies no torque~\cite{*[{}] [{, Sec.~3.5.}] Doyle07}.

The intriguing experimental findings have triggered a broad theoretical
search for the spontaneous rotation's physical origins. Although neoclassical
(collisional transport) effects have been considered \cite{Chang08, Rogister02, *Singh06, *Daybelge09},
extensive experimental evidence indicates that turbulence dominates
momentum transport \cite{deGrassie09rev,ScottSD90, *LeeWD03, *Kallenbach91, *Rice04nf}.
Numerical efforts have investigated turbulent momentum transport in
the core, both linearly \cite{Dominguez93, *Peeters07, *Peeters09, *Peeters09cor, *Kluy09, Camenen09, *Camenen09pop}
and nonlinearly \cite{Garbet02, *Waltz07, *Holod08, *Terry09, *Wang09, *Wang11}.
Models for intrinsic rotation, also primarily core-focused, have treated
quasilinear approximations \cite{Dominguez93, Shaing01, *Yoon10, *Singh11, Hahm07, Coppi02, *Gurcan07mom, Camenen09, McDevitt09, *McDevitt09pop, *McDevitt09dA},
effects of inhomogeneity of the confining magnetic field $\bv$ \cite{Hahm07,Camenen09},
nonresonant correlations between the fluctuating radial $\exb$ drift
$\ver$ and parallel velocity $\vlf$ \cite{Diamond08, *Diamond09, *Gurcan10mom_int},
and Stringer spin-up type effects \cite{Wobig95, *Peeters98, *Rozhansky10}.
Some scrape-off-layer (SOL) effects have been entertained \cite{Rozhansky94, *Chankin96, *Rozhansky08, LaBombard04},
but without systematic consideration of the confined plasma's response.

Momentum transport in the tokamak edge presents particular challenges
for theory. The turbulence is strong, with statistics very different
from quasilinear estimates \cite{Wakatani84, *Scott05dwvsbm}. It
is also strongly anisotropic, with parallel fluctuation length $\Ll$
two orders of magnitude larger than the radial length scale $\lp$
characterizing\emph{ }toroidal velocity and equilibrium plasma variation
\cite{Endler99, *Bleuel02, Putterich09, deGrassie09}. Since parallel
fluctuation gradients $\kl\sim1/\Ll$ and the corresponding forces
are accordingly weak, turbulently-accelerated $\vlf$ and the resulting
nondiffusive effects \cite{Coppi02,Gurcan07mom,Diamond08,McDevitt09,Gurcan10mom_int}
are smaller than simple diffusive momentum transport by $\kl\lp\ll1$
for realistic edge parameters. Most of these effects are further
reduced, actually proportional to a {}``symmetry-breaking'' $\langle\kl\rangle\ll\langle\kl^{2}\rangle^{1/2}\sim1/\Ll$
\cite{Peeters05,McDevitt09}. Since $\bv$ varies on the scale length
of the major radius $\Ro$, the resulting momentum transport effects
scale relative to simple diffusion as $\lp/\Ro\ll1$ in the edge \cite{Hahm07,Camenen09}.
 Further, the interaction of edge and SOL makes the problem inherently
radially nonlocal. For example,  the amplitude of the (unnormalized)
turbulent fluctuating potential decreases in the radial direction
on a short length scale $\lf\sim\lp$ \cite{Ritz90, *Endler95, *Moyer97, *Moyer99, *Silva04, *LaBombard05emagnetic, *Horacek10}.
Given these experimental facts, the present work analyzes a simplified,
purely-diffusive kinetic transport model, setting parallel acceleration
identically to zero but retaining a model edge and SOL, passing-ion
drift orbit excursions, and spatial variation of the diffusivity,
finding differential transport of co- and counter-current ions to
cause residual stress and consequent intrinsic rotation levels similar
to those seen in experiment.

Analysis begins with a model axisymmetric drift-kinetic transport
equation for the ions (c.\,f.~\cite{Catto94})\begin{equation}
\pt\fie+\vn\py\fie-\drat\vn^{2}\left(\sin\yy\right)\px\fie-\dy\left(\yy\right)\px\left(e^{-\xx}\px\fie\right)=0,\label{eq:kin_eqn_norm}\end{equation}
in which $\fie(\xx,\yy,\vn,t)$ is the ion parallel distribution function
normalized to pedestal-top ion density over thermal speed $\nni\ped/\vti\ped$
and $\vn$ is the parallel velocity normalized to $\vti\ped$, positive
for co-current motion. The geometry is simple slab plus geodesic curvature
drift, with uniform poloidal $\bpo$, toroidal $\bto$, and total
$\bom$ magnetic field strength. The radial position $\xx$, poloidal
position $\yy$, and time $t$ are respectively normalized to $\lf$,
the minor radius $\ro$, and the ion transit time $\ro\bom/\bpo\vti\ped$.
The effects of nonaxisymmetric fluctuations are modeled with an inhomogeneous
turbulent diffusivity, normalized to $\lf^{2}\bpo\vti\ped/\ro\bom$
and assumed separable, with arbitrary poloidal dependence $\dy(\yy)$,
decaying exponentially in $\xx$. The dimensionless parameter $\drat\doteq\qq\ri\ped/\lf$,
with $\qq\doteq\ro\bto/\Ro\bpo$ the safety factor and $\ri$ the
thermal ion gyroradius, indicates the passing-ion orbit width relative
to the radial turbulence inhomogeneity. $\drat$ takes values around
$1/4$ for typical ASDEX-Upgrade (AUG) H-mode parameters \cite{AUG_params11}.
Collisions are neglected, a reasonable approximation if pedestal-top
ions escape without experiencing a collision, roughly for $\nii\tc<1$,
with $\nii$ the velocity-dependent ion collision rate and $\tc$
the pedestal ion stored energy over the ion heat flux. For thermal
pedestal-top ions, $\nii\tc$ takes values around 1 for typical H-modes
in AUG, JET, and DIII-D, thus superthermal pedestal-top ions tend
to exit the plasma collisionlessly while subthermal ones do not.
Boundary conditions are $\fie(-\infty,\yy)\to\fio(\vn)$, $\fie(\infty,\yy)\to0$,
$\fie(\xx<0,\yo)=\fie(\xx<0,\yo+2\pi)$, $\fie(\xx>0,\yo,\vn>0)=0$,
and $\fie(x>0,\yo+2\pi,\vn<0)=0$, with $\yo$ the poloidal X-point
angle. Eq.~\eqref{eq:kin_eqn_norm} is invariant to a rigid toroidal
rotation $\vrig$, normalized to $\vti\bto/\bom$. Although the relevant
general theorem \cite{Scott10mom, *Brizard11} does not directly apply
to this non-Hamiltonian model, Eq.~\eqref{eq:kin_eqn_norm} does
trivially conserve a simplified toroidal angular momentum $\int(\vn+\vrig)\fie\,\dd\vn$,
as well as a density $\int\fie\,\dd\vn$ and energy $\int(1+\vn^{2}/2)\fie\,\dd\vn$.

Eq.~\eqref{eq:kin_eqn_norm} can be approximately analytically solved
$\vn$-by-$\vn$ in steady state for both large and small $\deff$,
results agreeing for $\deff\approx1$. The solution procedure is briefly
described here, details given in \cite{StoltzfusDueck11pop}. Since
$\vn$ appears only as a parameter, $\vn$-dependent variable transforms
can greatly simplify Eq.~\eqref{eq:kin_eqn_norm}. First, use new
spatial variables\begin{subequations}\label{eq:var_transforms}\begin{align}
\xb & \doteq\xx-\drat\vn\left(\cos\yy-\cos\yo\right),\label{eq:xbar_transform}\\
\yb & \doteq\dyo^{-1}\left(\vn\right)\int_{\yo}^{\yy}\dy\left(\yp\right)e^{-\drat\vn\left(\cos\yp-\cos\yo\right)}\dd\yp,\label{eq:ybar_transform}\end{align}
\end{subequations}with $\dyo(\vn)\doteq\int_{\yo}^{\yo+2\pi}\dy(\yp)\exp\boldlp-\drat\vn(\cos\yp-\cos\yo)\boldrp\dd\yp$.
Physically, $\xb$ is a drift-surface label and $\dyo$ an orbit-averaged
diffusivity. Next, for $\vn<0$, take $\yb\to1-\yb$. Finally, transform
from $\xb$ to $\uu\doteq e^{\xb/2}$, obtaining\begin{equation}
\pyb\fie=\left(\deff/4\right)\left(\pu^{2}\fie-\uu^{-1}\pu\fie\right)\label{eq:kin_eqn_transformed}\end{equation}
for $\fie(\uu,\yb,\vn)$, in which $\deff(\vn)\doteq\dyo(\vn)/|\vn|$.
Boundary conditions are now $\fie(0,\yb)=\fio(\vn)$, $\fie(\infty,\yb)\to0$,
$\fie(\uu<1,0)=\fie(\uu<1,1)$, and $\fie(\uu>1,0)=0$. The normalized
flux of particles with velocity $\vn$ through any closed poloidal
contour, $\fluxs(\vn)$, takes the simple form \begin{equation}
\fluxs\left(\vn\right)=-\frac{1}{2}\dyo\left(\vn\right)\uu^{-1}\int_{0}^{1}\pu\fie\,\dd\yb,\label{eq:flux_of_v}\end{equation}
evaluated at any constant $\uu\le1$.

\begin{figure}
\includegraphics[bb=0bp 0bp 488bp 291bp,clip,width=8.5cm]{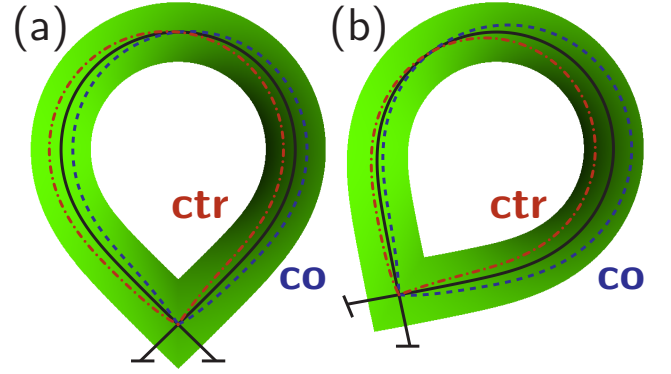}

\caption{\label{fig:drift_orbits}(color online). Co- and counter-current passing
ion drift orbits over turbulence, plotted for a straight-down (a)
and inboard (b) X-point. Darker shading indicates stronger diffusivity.
Co- (ctr-) orbits are displaced major-radially outward (inward) as
shown, regardless of the sign of $\bto$ or $\plascur$.}

\end{figure}

Eq.~\eqref{eq:kin_eqn_transformed} shows the original problem to
reduce to a one-parameter family of otherwise-identical differential
equations. Remarkably, the spatially-constant effective diffusivity
$\deff$ depends not only on the magnitude of $\vn$, but also on
its sign! Physically, as shown in Fig.~\ref{fig:drift_orbits}, this
results from the fact that co- (counter-) current ions' drift orbits
are displaced major-radially outwards (inwards). For the typical case
of turbulent diffusivities larger at the outboard, counter-current
ions experience larger orbit-averaged diffusivities. Preferentially
exhausting counter-current ions represents a co-current residual stress,
although momentum transport at any given spatial point is purely diffusive.

In solving Eq.~\eqref{eq:kin_eqn_transformed}, a Laplace transform
approach similar to \cite{*[{}] [{, App.~A.1.}] Farnell04} yielded
the exact Green's function,\begin{equation}
\gf\left(\uu,\sx,\yg\right)=\frac{2\uu}{\deff\yg}\exp\left(-\frac{\sx^{2}+\uu^{2}}{\deff\yg}\right)I_{1}\left(\frac{2\uu\sx}{\deff\yg}\right),\label{eq:greens_func}\end{equation}
in terms of which the solution may be written as\begin{equation}
\fie\left(\uu,\yb\right)=\fio e^{-\uu^{2}/\deff\yb}+\int_{0}^{1}\fie\left(\sx,0\right)\gf\left(\uu,\sx,\yb\right)\dd\sx.\label{eq:fie_of_greens_func}\end{equation}
A first-order iterative approximation $\fif$, obtained using $\fie(\sx,0)\to\fio\exp(-\sx^{2}/\deff)$
in Eq.~\eqref{eq:fie_of_greens_func}, was demonstrated to yield
an approximate normalized flux $\fluxs/\fio|\vn|$ with absolute error
strictly less than $\min\boldlp0.58\deff^{1/2}(1-e^{-1/\deff}),0.75/\deff^{3/2}\boldrp$,
tight bounds for large $\deff$. For small $\deff$, a two-region
solution may be used, representing $\fie$ with a Fourier series for
$\uu<1$ (edge) and Laplace transforming for $\uu>1$ (SOL), requiring
continuity in $\fie$ and $\pu\fie$ at $\uu=1$, except possibly
at the single point $\uu=1$, $\yb=0$. The resulting edge and SOL
ODEs possess explicit solutions in terms of modified Bessel functions.
Slightly generalizing \cite{Oldham72}, continuity at the LCFS then
requires the edge solution to satisfy\begin{equation}
\fie\approx-\frac{1}{2}\deff^{1/2}\frac{1}{\sqrt{\pi}}\int_{0}^{\yb}\frac{\pu\fie\,\dd\yp}{\sqrt{\yb-\yp}}-\frac{1}{8}\deff\int_{0}^{\yb}\pu\fie\,\dd\yp\label{eq:small_D_LCFS_condn}\end{equation}
at $\uu=1$. The resulting dense matrix for the Fourier coefficients
has been solved numerically at various $\deff$, retaining 10000 modes
in $\yb$. 

\begin{figure}
\includegraphics[bb=0bp 0bp 260bp 84bp,clip,width=8.5cm]{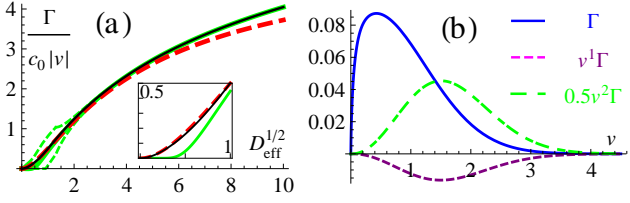}

\caption{\label{fig:flux_of_v}(color online). (a) normalized flux as a function
of the $\vn$-dependent effective diffusivity, with uniform approximation
(thin solid black), small-$\deff$ approximation (thick dashed red),
large-$\deff$ approximation (thick solid green), and large-$\deff$
error bounds (thin dashed green). (b) speed distribution of particle,
momentum and parallel heat fluxes, assuming a Maxwellian at the inner
boundary.}

\end{figure}

The two approximate solutions,\begin{equation}
\frac{\fluxs}{\fio\left|\vn\right|}\approx\begin{cases}
\deff/\left(1+\aone\deff^{1/2}+\at\deff\right), & \deff\lesssim1\\
-\frac{1}{2}\deff\int_{0}^{1}\pu|_{\uu=1}\left(\fif/\fio\right)\,\dd\yb, & \deff\gtrsim1\end{cases},\label{eq:flux_soln_limits}\end{equation}
are well-approximated for all $\deff$ by \begin{equation}
\frac{\fluxs}{\fio\left|\vn\right|}\approx\frac{1}{4}\ln\left(1+\sum_{j=2}^{8}c_{j}\deff^{j/2}\right)\approx\ln\left(1+\frac{\deff}{e^{\gamma}}\right),\label{eq:flux_soln_uniform_appns}\end{equation}
with $\aone=0.8224$, $\at=0.1763$, $c_{2}=4$, $c_{3}=-4\aone$,
$c_{4}=4\aone+e^{4/(1+\aone+\at)}-5-9e^{-4\gamma}$, $c_{5}=c_{7}=0$,
$c_{6}=8e^{-4\gamma}$, $c_{8}=e^{-4\gamma}$, and Euler's constant
$\gamma\approx0.5772$. The second approximation is used for the simplified
explicit forms in Eqs.~\eqref{eq:moment_fluxes}--\eqref{eq:int_rotn}
and corresponding plots, the first for all other plots. The results
of Eqs.~\eqref{eq:flux_soln_limits} and~\eqref{eq:flux_soln_uniform_appns}
are plotted in Fig.~\ref{fig:flux_of_v}(a), along with the large-$\deff$
error bounds.

Explicit forms for normalized particle, momentum, and parallel heat
fluxes may now be obtained for any specified $\fio(\vn)$ and $\dy(\yy)$.
Assuming a Maxwellian $\fio(\vn)=e^{-\vn^{2}/2}/(2\pi)^{1/2}$ and
simple ballooning form $\dy(\yy)=\dnaught(1+\fb\cos\yy)$, thus\begin{equation}
\deff\left(\vn\right)=2\pi\dnaught e^{\drat\vn\cos\yo}\left[I_{0}\left(\drat\vn\right)-\fb I_{1}\left(\drat\vn\right)\right]/\left|\vn\right|,\label{eq:dyo_for_sinusoidal_D}\end{equation}
the relevant flux moments may be reasonably approximated for small
$\drat$ as \begin{subequations}\label{eq:moment_fluxes}\begin{align}
\fluxp & \doteq\int_{-\infty}^{\infty}\fluxs\,\dd\vn\hphantom{v^{2}}\approx\sqrt{\frac{2}{\pi}}\logf_{1}\!,\label{eq:dens_flux}\\
\fluxm\,\, & \doteq\int_{-\infty}^{\infty}\vn\fluxs\,\dd\vn\approx8\drat\sqrt{\frac{2}{\pi}}\left(\cos\yo-\frac{\fb}{2}\right)\left(\logf_{3}-\logf_{5}\right)\!,\label{eq:mom_flux}\\
\fluxh & \doteq\int_{-\infty}^{\infty}\frac{\vn^{2}}{2}\fluxs\,\dd\vn\approx\sqrt{\frac{2}{\pi}}\logf_{3}\!,\label{eq:heat_flux}\end{align}
\end{subequations}in which $\logf_{p}(\dnaught)\doteq\ln(1+2\pi\dnaught/e^{\gamma}p^{1/2})$.
The integrands are plotted in Fig.~\ref{fig:flux_of_v}(b), summed
over the sign of $\vn$. (All plots use representative AUG H-mode
values $\dnaught=0.033$, $\fb=0.8$, $\yo=-5\pi/8$, and $\drat=0.28$.) 

\begin{figure}
\includegraphics[clip,width=8.5cm]{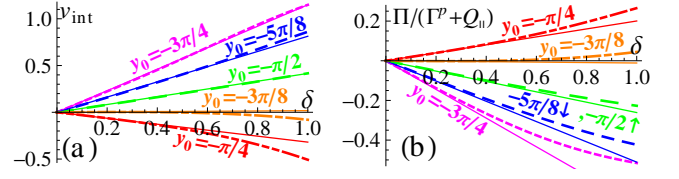}

\caption{\label{fig:rotation_results}(color online). Normalized intrinsic
rotation velocity $\vint$ (a) and unbalanced NBI fraction $\fluxm/(\fluxp+\fluxh)$
(b), plotted as functions of drift orbit width $\drat$ for several
values of poloidal X-point angle $\yo$, with numerical integrals
(thick dashed) and analytical approximations (thin solid).}

\end{figure}

The total momentum flux, incorporating $\vrig$, is just $\vrig\fluxp+\fluxm$.
Since toroidal rotation damping is very weak \cite{Connor87}, a vanishing
momentum input implies that momentum flux must also vanish, resulting
in the pedestal-top intrinsic rotation rate\begin{equation}
\vint=-\frac{\fluxm}{\fluxp}\approx8\drat\left(\fb/2-\cos\yo\right)\frac{\logf_{3}-\logf_{5}}{\logf_{1}},\label{eq:int_rotn}\end{equation}
plotted in Fig.~\ref{fig:rotation_results}(a). Alternatively, one
may balance an applied NBI torque with the outward momentum flux resulting
from the NBI-driven ion heat flux. For zero pedestal-top toroidal
rotation, $\vrig=0$, one must set the unbalanced NBI fraction $\funb\doteq(P_{\text{NBI}}^{\text{co}}-P_{\text{NBI}}^{\text{ctr}})/P_{\text{NBI}}$
to\begin{equation}
\funb=\frac{\fconvi}{2\fnbi}\frac{\bto}{\bom}\frac{\vbeam}{\vti\ped}\frac{\fluxm}{\fluxp+\fluxh},\label{eq:unbalanced_nbi_frac}\end{equation}
in which $\fnbi$ is the fraction of heating by NBI, $\fconvi$ is
the fraction of heat transported by ions, and $\vbeam$ is the beam
ion velocity. The ratio $\fluxm/(\fluxp+\fluxh)$ is plotted in Fig.~\ref{fig:rotation_results}(b).
Since $\vbeam/\vti\ped$ is typically large, $\funb$ may be a significant
fraction of unity, as observed by \cite{Solomon07}.

Several features of this solution deserve comment. First, the steady-state
results given in Eqs.~\eqref{eq:int_rotn}--\eqref{eq:unbalanced_nbi_frac}
are due to a balance between large momentum transport terms {[}Fig.~\ref{fig:flux_of_v}(b){]},
thus are robust. Relaxation time to the edge intrinsic rotation profile
should occur roughly on an ion transport time through the pedestal,
$\sim\tc$. The $\vn$-asymmetric diffusion $\fluxm$ is independent
of the toroidal velocity and its radial gradient, thus $-\fluxm$
represents a residual stress. For typical experimental parameters,
it acts in the co-current direction with experimentally-relevant magnitude.
The dimensional intrinsic rotation prediction, given here for small
$\dnaught$, is \begin{equation}
\vintd\approx1.04\frac{\bto}{\bom}\left(\frac{\fb}{2}-\cos\yo\right)\frac{\qq\ri\ped}{\lf}\vti\ped\propto\frac{\Ti\ped}{\bpo}.\label{eq:int_rotn_diml}\end{equation}
 The $1/\bpo$ dependence corresponds to experimentally-observed
$1/\plascur$ scalings \cite{deGrassie09rev,Rice07}, while proportionality
to $\Ti\ped$ provides an alternative explanation for recent observations
\cite{deGrassie09,Rice11}. Co-current spin-up at the L-H transition
\cite{deGrassie09rev,Rice07} is expected due to the increase in $\Ti\ped$
and probable decrease in $\lf$. The predicted dependence on X-point
poloidal angle has yet to be experimentally tested. The physics presented
here may also have implications for internal transport barrier (ITB)
rotation: for outboard-ballooning and radially-increasing diffusivity
(as outside an ITB), the asymmetric diffusivity results in a counter-current
core rotation increment, consistent with \cite{Rice01,Rice08}.

Simplifications used in this model must be kept in mind. The presented
calculations omitted both the $\nabla B$ drift and the radial electric
field $\Er$, outside the latter's contribution to $\vrig$. While
the $\nabla B$ drift has little effect, a uniform uncanceled poloidal
$\exb$ drift of magnitude approaching $\vti\bpo/\bom$ can represent
a nonnegligible co- (counter-) residual stress for outwards (inwards)
$\Er$, a transport effect due to a shifted relation between $\dyo$
and $\deff$ \cite{StoltzfusDueck11pop}, possibly connected with
observations of a favorable/unfavorable X-point dependence of L-mode
toroidal rotation \cite{LaBombard04,Rice08}.  Treatment of sheared
$\Er$ effects or electrostatic confinement \cite{Catto78adiab} would
require nontrivial extensions to the theory, as would retention of
collisions. Direct collisional effects on the rotation-driving flux
$\fluxm$ may often be small, since $\fluxm$ results dominantly from
ions that are slightly superthermal at the pedestal top {[}Fig.~\ref{fig:flux_of_v}(b){]},
thus very superthermal at the LCFS, with an accordingly low collision
rate. However, lower-energy ions may affect both $\Er$ and the rotation
saturation $\vrig\fluxp$. Finally, recall that the turbulence parameters
are taken as an input to the present model, not calculated self-consistently.

In summary, radial displacements of passing-ion orbits and typical
tokamak-edge turbulence inhomogeneity are shown to result in orbit-averaged
diffusivities that depend on the sign of $\vl$. Even in the absence
of nondiffusive effects, this results in residual stress and corresponding
intrinsic rotation at experimentally-relevant levels. The rotation
is co-current for typical H-mode parameters and scales with $\Ti\ped/\bpo$,
in agreement with experimental observations.
\begin{acknowledgments}
Helpful discussions with A.~Chankin, G.~Hammett, P.~Helander,
J.~Krommes, K.~Lackner, O.~Maj, R.~McDermott, B.~Nold, T.~Pütterich,
C.~Rost, P.~Schneider, and B.~Scott, and an Alexander von Humboldt
Foundation research fellowship are gratefully acknowledged.
\end{acknowledgments}
\bibliographystyle{apsrev4-1}
\bibliography{tjsd}

\end{document}